# Small universal E-taster for rapid comparative analysis of liquids

Tatiana Yakhno[1*], Alexander Pakhomov[1], Anatoly Sanin[1], Vyacheslav Kazakov[1], Ruben Ginoian[2], Vladimir Yakhno[1]

[1]Institute of Applied Physics RAS, 46 Ulyanov Street, Nizhny Novgorod, Russia

[2]Nizhny Novgorod State Agricultural Academy, 97 Gagarin Ave, Nizhny Novgorod, Russia

(T.Y.)* e-mail: yakhta13@gmail.com

**Abstract**

An original method and a device for obtaining "fingerprints" of liquids are proposed. According to this method, a "fingerprint" is a dynamic parameter – an oscillogram of the electrical impedance of a quartz sensor on which a drop of the studied liquid dries. The design of the device and the principle of its operation are described. The method can be used for a quick and inexpensive assessment of the degree of compliance of a liquid with the standard. Moreover, the degree of the difference between liquids of different compositions is expressed quantitatively. The device is a compact console for a laptop. The method is implemented in the range of room temperatures and humidity. The work of the E-taster is demonstrated by comparing the "fingerprints" of spirits, wines of different types and brands and solutions of milk powder of different concentrations.

## 1. Introduction

Nowadays, a huge variety of touch-sensitive devices such as "E-tongue" [1,2] or "E-nose" [3-5] are offered for quick and inexpensive assessment of the quality of liquid food products produced or stored in a warehouse – juices, milk, beer, wine, strong alcohol. Evaluation of the quality of the tested products using such devices is, as a rule, based on the qualitative/quantitative determination of one or more components of the product, clearly indicating its deterioration or adulteration. For this purpose, coatings that specifically bind the chemicals of interest in the gas or liquid phase are applied onto the surface of a sensor (group of sensors) and their mass is determined relative to the control using QCM [3-7]. Cleaning or replacing sensor coatings complicates practical application of these devices, so a rare "E-nose" goes beyond the boundaries of research laboratories.

Another, holistic approach is being developed to assess the authenticity of a particular food product. It is based on comparing "fingerprints" of non-targeting spectroscopy and spectrometry of reference and analyzed products using known analytical methods (UV-vis[1], NIR[2], FT-MIR[3], [1]H

---
[1] Ultraviolet–visible spectroscopy;
[2] Near Infrared Spectropy
[3] Fourier-transform mid infrared spectroscopy

NMR[4], C[13] NMR[5], etc.) using multivariate statistical analysis [8]. This approach can be used for determining the geographical origin of the product, control of its component composition and is aimed at confirming its authenticity or identifying fraud. The authors of the review [8] suggest ways to harmonize and standardize methods for comparing "fingerprints" as a prerequisite for analyzing the authenticity of food products within the framework of official control.

A method of medical diagnostics based on the morphological features of microstructures in dried drops of human biological fluids was widely used at the beginning of the XXI-st century [9]. It was shown that, under room conditions, the pattern of the droplets dried on glass is reproducible and is based on the chemical composition of biological fluids, which can be used as an additional diagnostic criterion. The only factor that reduced the express advantages of the method was the need for a long (at least 2 days) preliminary drying of these drops under room conditions to obtain a clear image. This methodological approach has been currently confirmed and further developed [10-12]. It was shown that the processing of video images of drops and the decision-making system can be implemented using Machine Learning Analysis based on both Linear Discriminant Analysis and Principal Component Analysis [13]. In addition to medical diagnostics, the prospects of analyzing microstructures in dried drops of multicomponent liquids for assessing their quality are also shown for a number of technical liquids [14,15], dairy products [16], and alcoholic beverages [17,18]. Interestingly, to organize the structures in initially homogeneous (at the micro level) dried drops of alcoholic drinks, crystalline NaCl was added to liquid tequila [17], and whiskey were diluted with water [18].

The goal of this work is to present a new methodological approach and a device based on it, suitable for rapid comparative analysis of liquid food products. It combines the informativeness of the dynamic processes of self-organization of drying drops and the registration of this dynamics using a sensor device with a quartz resonator on which the drop dries. As an integral characteristic of the liquid, the "fingerprint", i.e., an oscillogram of the change in the electrical impedance of quartz is used, which reflects the dynamics of the complex mechanical properties of the droplet during drying. This approach does not require visually distinguishable microstructures in dried drops and is also suitable for work with liquids, the drops of which, when dried, form a skin layer or gelatinize (juices, wines). The method does not require any preliminary sample preparation and coating of the sensor surface; the recording time (15-20 min) corresponds to the drying time of a droplet with a volume of 3 µl under room conditions. The dynamics of the mechanical properties of drying drops of the considered liquids is compared automatically, with the results of the obtained quantitative differences displayed on the monitor.

---

[4] Proton nuclear magnetic resonance
[5] Carbon-13 nuclear magnetic resonance

## 2. Theory

### 2.1. Technical Description

The detailed technical description of the method was presented in [19]. Here we will demonstrate only some basic points supported by results of the previous studies. The functional diagram of the device is shown in Fig. 1.

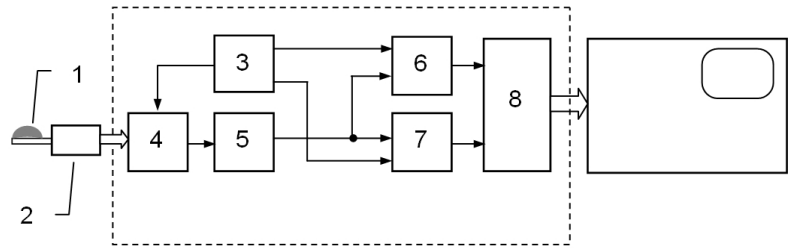

Fig. 1. Functional diagram of the device. 1 – drop; 2 – sensor; 3 – generator; 4 – bridge; 5 – amplifier; 6 and 7 – phase detector; 8 – microcontroller. The block on the right is a laptop.

This device allows recording the dynamics of the mechanical properties of the droplets drying on the surface of a quartz resonator using acoustic impedance measurements. The recorded value is called "Acoustomechanical impedance" (AMI) [19]. AMI measurements are based on the dependence of the electrical characteristics of a resonator on the physical properties of a liquid. This dependence is widely used in the studies of the properties of gases and liquids by means of electro-acoustic resonators QCM [20-23]. The measured electrical characteristics of a resonator are, as a rule, its resonance frequency and Q-factor, which change when the resonator contacts the object under study. This property is usually used as the basis for determining the physical properties of a liquid such as viscosity, density, concentration of sought substance, and so on. A radical difference of our method from the known ones is that we use the temporal dependence of the AMI of the drying drop as the informative parameter [19]. At the same time, the resonator oscillation frequency of 60 kHz is kept constant during measurements. A drop of liquid under study is placed at the end of the plate (Fig. 2), i.e., where the oscillatory velocity amplitude of the surface is approximately constant. This end of the plate is for the operation. Part of the area of the electrodes at the operating end of the plate was removed to place the drop directly on the surface of a quartz crystal (Fig. 2). Under such conditions, the drop is an acoustic (mechanical) load of the resonator for shear oscillations.

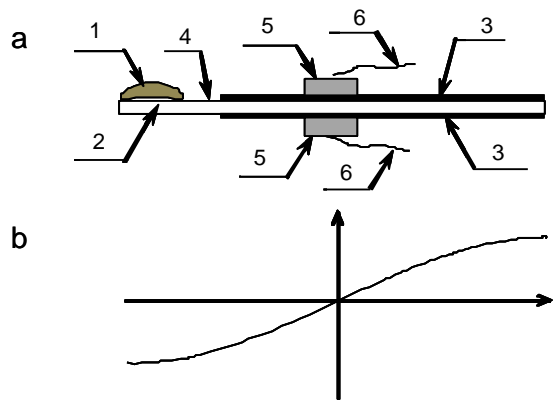

Fig. 2. Sensor device: (a) 1 – drop, 2 – quartz resonator; 3 – metallization; 4 – surface of quartz without metallization; 5 – conductive rubber; 6 – conductors. (b) The distribution of shear vibrational velocity along the resonator length.

## 2.2. Signal Registration

Drying of the droplet is accompanied by a regular change in its physical characteristics, which is reflected in the corresponding changes in the electrical parameters of the quartz resonator. The module and phase of the electrical impedance of quartz or the real and imaginary parts of the quartz impedance can be the recorded parameters. All these parameters change in time as the droplet is drying. The recording of any of them is, in fact, an oscillogram of the test liquid drying. During the droplet drying and transition to the solid state, the mechanical interaction of the droplet residues with the sensor increases and manifests itself in an increase in the signal of the module and a decrease in the signal of the AMI phase. Upon reaching the ultimate mechanical stress (critical point, a), the signal decreases as a result of microcracks in the dried material and its partial delamination [24] (Fig. 3). This is the basic pattern of the change in the module and phase of AMI for all aqueous solutions. However, changes in the geometry and amplitude of the signals depend on the composition, concentration, and dispersion of each liquid, which allows the differences between the liquids to be assessed quantitatively.

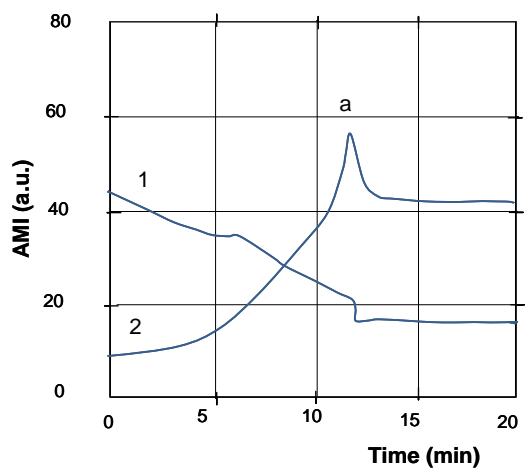

Fig. 3. Change in the amplitude (1) and phase (2) of AMI during drying of a drop on the sensor surface.

**2.3. Signal processing**

Simultaneous registration of two parameters, namely, the real Re (Z) and imaginary Im (Z) parts of the electric impedance Z of the sensor allows representing the results in the form of a hodograph — a curve on the plane in the coordinates of the real and imaginary parts of the electric impedance. Such a waveform representation of changes in the electrical impedance of quartz is also a "fingerprint" of the liquid.

The quantitative difference between the liquids was found using two approaches: parametrization of the shape of the amplitude curves and comparison of the relative positions of the hodographs of the compared liquids on a complex plane.

**2.3.1. Amplitude curve parametrization (Shape Indices)**

An indispensable stage of processing the curves is choosing their informative areas. At this stage of work, we analyzed the shape of the amplitude component of the AMI in the area of its peak (Fig. 4).

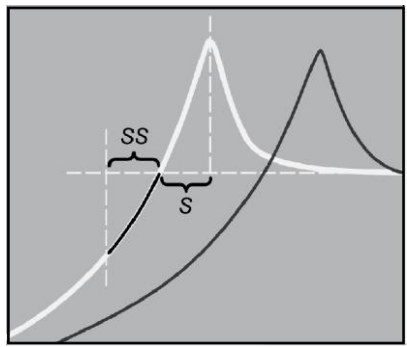

Fig. 4. Finding reference points and determining design parameters characterizing the shape of the amplitude curve of a drying drop[6].

We chose the combinations of parameters yielding the largest differences when two particular groups of liquids were compared. Figure 4 shows the reference points automatically chosen on the amplitude curve to calculate the pre-assigned parameters characterizing particular features of the curve shape. The calculated values of the parameters were displayed on the monitor. After the drop dried up, the calculation of the resulting curve was almost instantaneous. The compared results were presented in the form of an arithmetic mean with a standard deviation (M±2σ) on the plane in the coordinates of different Shape Indices (SI). The results of parametrization of the impedance module curves for juices of different brands purchased in the distribution network: 4 apple, 2 grape and 3 tomato brands, as well as grape drink are presented in Fig. 5. In the SI coordinates 4 and 6 on the plane (Fig. 5a), the juices are grouped according to the raw materials (apples, grapes, tomatoes, grape drink). In the SI coordinates 1 and 6 (Fig. 5b), the array of points 1 in Fig. 5a is divided into four varieties of apple juice of different brands. In the SI coordinates 1 and 4 (Fig. 5c) the array of points 4 in Fig. 5a is divided into three varieties of tomato juice.

---

[6] Description of Algorithms for SI calculation can be found in Appendix to the paper [27].

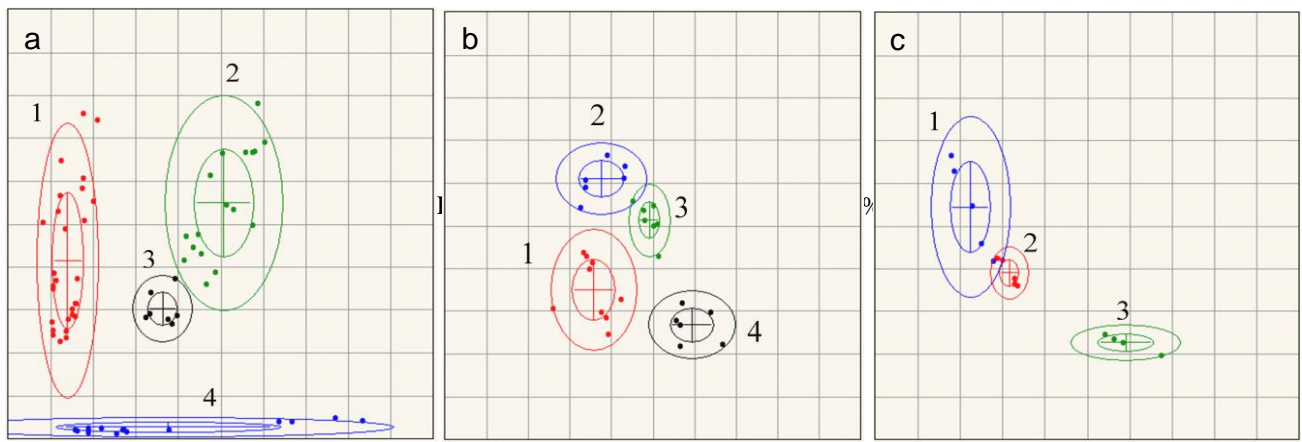

GOLD, J-7, Nico, and My Family; 2 – 100% grape juice of two brands: Rich and I; 3 – grape drink J-7; 4 – 100% tomato juice with pulp of three brands: J-7, Nico, and I. (b) Plane in SI coordinates 1 and 6 (M ± 2σ). Apple juice of different brands: 1 – GOLD; 2 – My Family; 3 – J-7; 4 – Nico. (c) Plane in SI coordinates 1 and 4 (M ± 2σ). Tomato juice of different brands: 1 – I, 2 – J-7 and 3 – Nico.

Thus, the shape of the curve (oscillograms) of the quartz impedance module upon drying of liquid droplets depends on the composition of these liquids and their dispersion.

The method also proved to be informative for medical [25,26] and veterinary [27] diagnostics, as well as for registration of liquid exposure to external factors of physical nature [28-30]. The creation of artificial protein-salt composites allowed revealing the specific contribution of individual components to the mechanical properties of liquid and drying solutions [31-35]. Thus, consideration of the structural evolution of drying drops on the basis of the physical chemistry of solutions, physics of polymers, mechanics, and materials science makes it possible to explain the phenomenology of the process in terms of well-known physical phenomena [24].

We had a successful collaboration with a winery on parallel assessment of the quality of wines [36]. All samples were analyzed at the Scientific Center "Wine-Making" with the FTIR express-analyzer WineScan Flex (FOSS, Denmark) in order to determine the main physicochemical parameters (ethanol, titratable and volatile acidity, reducing sugars, reduced extract, pH, $SO_2$, phenolic substances). We compared four groups of samples: red dry wine (22 brands), white dry wine (15 brands), red semi-dry wine (6 brands) and white semi-dry wine (5 brands). We proved experimentally that the AMI curve is a passport characteristic of wine. Two approaches were used for parametrization of the AMI curves based on their geometrical features. One of them was calculation of Shape Indices, and the other was wine recognition by learning self-organizing neural networks. In our subsequent joint work [37] we traced the correlation between wine composition and SI, on the one hand, and between the composition of the same wine and the estimates of the panel of regional tasting experts, on the other hand. In quantitative terms, our technology shows higher sensitivity to the composition of wines than the assessment of tasters.

## 2.3.2. Parametrization of results using both the real and imaginary parts of the sensor electrical impedance

To improve the operational and informational characteristics, the implementation of the described method has been changed substantially: the dimensions of the device have been reduced, a modern element base has been used, the sensor module with an USB connector has been made removable to allow gentle cleaning of the sensor between measurements. A new interface has been developed: both the real and imaginary parts of electric impedance of the sensor in the form of hodographs on a complex plane are used for analysis (Fig. 6).

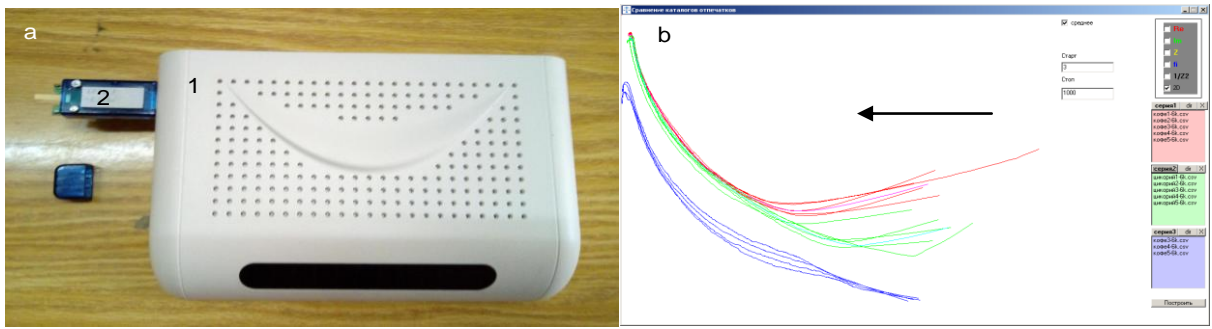

Fig. 6. (a) Photo of a new version of the device: 1 – housing, 2 – detachable sensor module. Housing dimensions: 21x12x4 cm, weight 0.2 kg; (b) software interface displayed on the monitor. Complex plane in Re/Im coordinates (a.u.) with hodographs of three coffee drinks of different brands of the same concentration. On the right is a list of files displayed on the monitor. The arrow indicates the direction of the drying process.

The electronic unit of the device (the dashed rectangle in Fig. 1) has a built-in microcontroller and is designed to convert the signal from the quartz resonator to voltages corresponding to the real and imaginary parts of its electrical impedance, which change during the drying process. The output voltage can be converted into the mechanical characteristics of a drying drop; however, for a comparative analysis of liquids, such a conversion is not necessary. The data is transferred via USB to a laptop, where it is visualized, processed and stored.

The software has a modular structure and includes 5 applications:

1. Recorder of the dynamics of drop drying

2. Viewer of saved data files

3. Application for searching the "fingerprint" of the liquid in the database (hodographs)

4. Application for comparing fingerprint catalogs

5. Application for searching the "fingerprint" of the liquid in the database (support vector method).

There are two modes of operation – the mode of creating the database and the mode of recording the "fingerprint" of the liquid for comparison. All applications work with a unified data format.

The viewer of stored data files permits displaying the stored data and make a preliminary comparison of the "fingerprints" of different liquids visually, with the "fingerprints" of one type of liquid are displayed in one color. The program for comparing the catalogs of dynamic portraits calculates the distances between the hodographs related to a series of repeated measurements of the drying process of one liquid and between the hodographs related to two different liquids. This allows comparing the catalogs of "fingerprints" of two liquids inside the database to reveal statistical differences between them (1). The hodograph Difference Index (DI) is calculated as the average distance between the points of the corresponding samples of two hodographs (Fig. 7).

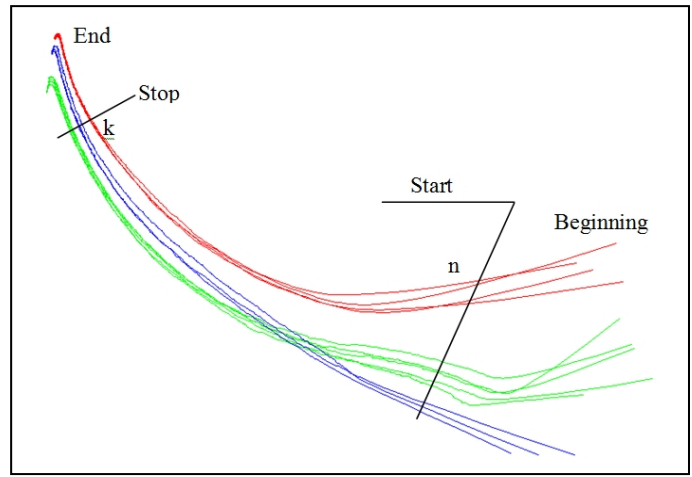

Fig. 7. An example of calculating the DI of two hodographs as the average distance between the points of the corresponding samples: n is the starting point of reference; k is the end point of reference.

The DI of the hodographs is calculated by the formula

$$DI = \frac{\sum_{i=n}^{k} \sqrt{(x_{i1} - x_{i2})^2 + (y_{i1} - y_{i2})^2}}{k - n} \quad , \tag{1}$$

where $x_{i1}$ is the x coordinate of the first hodograph, $x_{i2}$ is the x coordinate of the second hodograph, $y_{i1}$ is the y coordinate of the first hodograph, $y_{i2}$ is the y coordinate of the second hodograph, n is the starting point of calculation, and k is the end point of calculation. When two identical datasets were compared, the DI was zero. The more differences between the hodographs, the larger the DI magnitude is. Using the automatically constructed hodographs, it is possible to estimate quantitatively the extent of the difference between the liquids of these two groups taking into account errors of the 1st and 2nd kind (level of intersection of the corresponding histograms, Fig. 8).

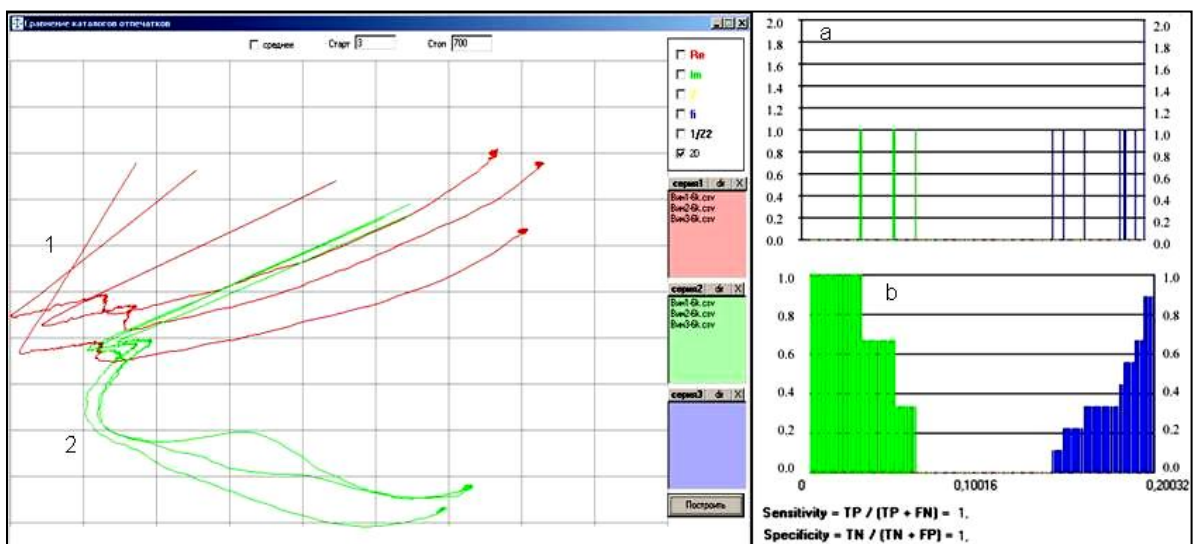

Fig. 8. Software interface. On the left are the hodographs of electrical impedance: 1 – port wine "777" (Russia) and 2 – vermouth "Martini Bianko" (Italy). Right: (a) Comparison of hodograph DI within each sample (errors of the first kind); (b) joint distribution of the hodograph DI of the compared wines (errors of the second kind) with the calculation of sensitivity and specificity [38].

On completion of the calculations, a list of liquids in the group sorted in ascending order of the averaged minimum distances and the "Result" window in which the liquid with the greatest similarity to the tested one is indicated (not shown) can be displayed on the monitor. This enables finding liquids in the group the hodographs of which are located closest to the tested one, and, therefore, having the greatest similarity with it.

3. **Experimental**

Students of higher educational institutions of Nizhny Novgorod took part in the tests of the new device. Whole milk powder, wine and hard drinks were bought in the city shops (Table 1). The experiments were carried out in laboratories at room temperature (20-23º C) and humidity (40-60%). Before each entry, current file name was set in the working window of the program. The registration software automatically tuned the frequency in the ± 1 Hz range for approximately 40 seconds. Upon completion of the automatic frequency tuning, a drop of the studied liquid with a volume of 3 μl was placed using a microdoser on the surface of the sensor, which served as a signal to start recording. Recording ended at the end of the time set by the operator. For each test liquid, 3-7 drops were measured. After each measurement, the remnants of the dried drop were carefully washed off the surface of the quartz plate with a cloth moistened with distilled water, wiped with a dry cloth and then with an alcohol wipe. After that, the next measurement was made. 25 samples were used in the study, including hard drinks, red and white wine, and milk (Table 1).

Table 1
List of the samples under study

| Hard Drinks | Wine |
|---|---|
| 1. Georgian brandy Eristavi 5* | 1. Semi-sweet red table wine Toro de Oro (Spain) |
| 2. Georgian brandy Eristavi 3* | 2. Port wine 777 (Russia) |
| 3. Russian brandy Staraya Krepost | 3. Vermouth Martini Bianco (Italy) |
| 4. Russian brandy Lezginka | 4. Semi-sweet grape natural red wine Khvanchkara (Georgia) |
| 5. Russian drink Three old men | |
| 6. Irish whiskey Jameson | 5. Dry white table wine Chardonnay (Fanagoria, Russia) |
| 7. Macallan Whiskey (Scotland) | 6. Dry white table wine Sauvignon (Fanagoria, Russia) |
| 8. Mordovian balm (Russia) | 7. Dry white table wine Aligote - Riesling of Fanagoria (Fanagoria, Russia) |
| 9. Brandy Stareishina (Russia) | |
| 10. Vodka Khlebnaya Sleza (Russia) | 8. Semi-dry red wine Pirosmani (Georgia) |
| | 9. Semi-dry red wine Pirosmani (Georgia) + sugar (3 g/100 ml) |
| | 10. Semi-dry red wine Pirosmani (Georgia) + sugar (6 g/100 ml) |
| | 11. Semi-dry red wine Pirosmani (Georgia) + sugar (9 g/100 ml) |

Milk

1. Whole powdered milk, 6 g/100 ml distilled water
2. Whole powdered milk, 8 g/100 ml distilled water
3. Whole powdered milk, 10 g/100 ml distilled water
4. Whole powdered milk, 14 g/100 ml distilled water

The results of the study depicted in the form of hodographs of electrical impedance of quartz on a complex plane demonstrate a wide variety of shapes depending on the type of drink (Fig. 9).

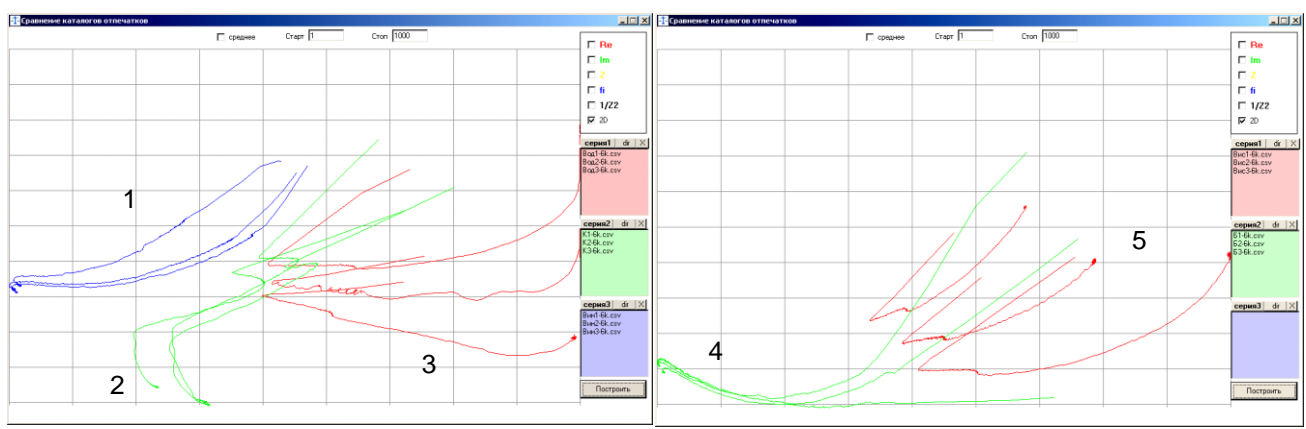

Fig. 9. Software interface. A variety of shapes of hodographs of liquids of different classes: 1 – dry red wine; 2 – cognac; 3 – whiskey; 4 – balm; 5 – vodka.

The recognition statistics – Sensitivity and Specificity – based on the calculation of the hodographs DI of the compared samples and errors of the 1st and 2nd kind are presented in Table 2.

Table 2

The obtained differences between the compared samples

| Compared Samples | Recognition statistics (EER) | |
| --- | --- | --- |
| | Sensitivity | Specificity |
| Hard drinks | | |
| 1. Georgian brandy Eristavi 5* & Georgian brandy Eristavi 3* | 0.932 | 0.989 |
| 2. Georgian brandy Eristavi 5* & Russian brandy Staraya Krepost | 0.969 | 0.996 |
| 3. Georgian brandy Eristavi 3* & Russian brandy Staraya Krepost | 0.203 | 0.987 |
| 4. Russian brandy Lezginka & Russian drink Three old men | 1.000 | 1.000 |
| 5. Russian brandy Lezginka & Irish whiskey Jameson | 1.000 | 1.000 |
| 6. Macallan Whiskey (Scotland) & Mordovian balm (Russia) | 1.000 | 1.000 |
| 7. Vodka Khlebnaya Sleza (Russia) & Brandy Stareishina (Russia) | 1.000 | 1.000 |
| Wine | | |
| 8. Semi-sweet red table wine Toro de Oro (Spain) & Port wine 777 (Russia) | 1.000 | 1.000 |
| 9. Port wine 777 (Russia) & Vermouth Martini Bianco (Italy) | 1.000 | 1.000 |
| 10. Semi-sweet red table wine Toro de Oro (Spain) & Semi-sweet grape natural red wine Khvanchkara (Georgia) | 1.000 | 1.000 |
| 11. Dry white table wine Chardonnay (Fanagoria, Russia) & Dry white table wine Sauvignon (Fanagoria, Russia) | 1.000 | 1.000 |
| 12. Dry white table wine Sauvignon (Fanagoria, Russia) & Dry white table wine Aligote - Riesling of Fanagoria (Russia) | 1.000 | 1.000 |
| 13. Semi-dry red wine Pirosmani (Georgia) & Semi-dry red wine Pirosmani (Georgia) + sugar (3 g/100 ml) | 1.000 | 1.000 |
| 14. Semi-dry red wine Pirosmani (Georgia) + sugar (3 g/100 ml) & Semi-dry red wine Pirosmani (Georgia) + sugar (6 g/100 ml) | 1.000 | 1.000 |
| 15. Semi-dry red wine Pirosmani (Georgia) + sugar (6 g/100 ml) & Semi-dry red wine Pirosmani (Georgia) + sugar (9 g/100 ml) | 1.000 | 1.000 |
| Milk | | |
| 16. Whole powdered milk, 6 g/100 ml distilled water & | | |
| Whole powdered milk, 8 g/100 ml distilled water | 0.992 | 0.851 |
| 17. Whole powdered milk, 6 g/100 ml distilled water & | | |
| Whole powdered milk, 10 g/100 ml distilled water | 1.000 | 1.000 |
| 18. Whole powdered milk, 10 g/100 ml distilled water & | | |
| Whole powdered milk, 14 g/100 ml distilled water | 1.000 | 1.000 |

According to the results of the tests (Table 2), the new implementation of the method demonstrated sensitivity sufficient for distinguishing drinks and a number of user advantages over the previous version. An unexpected result was low sensitivity of distinguishing between Georgian brandy Eristavi 3 * and Russian brandy Staraya Krepost (0.203), which indicates the similarity of these drinks.

## 4. Discussion

In this paper, we presented a new approach for rapid assessment of the similarity/difference of compared drinks – multicomponent liquids – without determining their composition. The information content of our method is based on the high sensitivity of the electrical characteristics of the oscillating sensor to the dynamics of the physical properties of a drop of the test liquid drying on its surface. As a result, each liquid acquires its own "dynamic portrait" (imprint), which reflects the component composition and dispersion of the liquid under given (room) conditions. In different implementations of the method, the fingerprint may be represented either by the shape of the amplitude curve (SI) or the difference index (DI) in the relative position of the hodographs on the complex plane, indicating the statistical differences between them – Sensitivity and Specificity. In either case the extend of the difference between the liquids is expressed quantitatively. The scientific novelty of our approach was confirmed by the patents [39-41]. Depending on the user need, the proposed device can be used to analyze any liquids (compare them with the corresponding standards). The only condition is that liquids must evaporate under room conditions. We believe that the considered quick and accurate evaluation of the similarity or difference of compared liquids is highly promising for preliminary rapid assessment of the quality of liquids, in particular, juices and wines, which can speed up and reduce the volume and cost of the quality assessment procedure using methods of standardized examination (Fig. 10). This will require creating specialized databases of reference samples collected in advance by the interested user. The coverage of possible falsifications with the use of this method will be much wider than that ensured by the control of individual indicators proposed by a number of researchers for specific samples [42,43]. If there is a sufficient number of databases of reference liquids, the closeness of a certain sample to one or another database will mean that it belongs to a particular class of liquids. This is a real way to liquid certification. The dynamic portrait (fingerprint) of a liquid could replace the barcode on product packaging, since the label on the packaging, strictly speaking, does not correlate with the quality of the liquid inside it and can also be falsified.

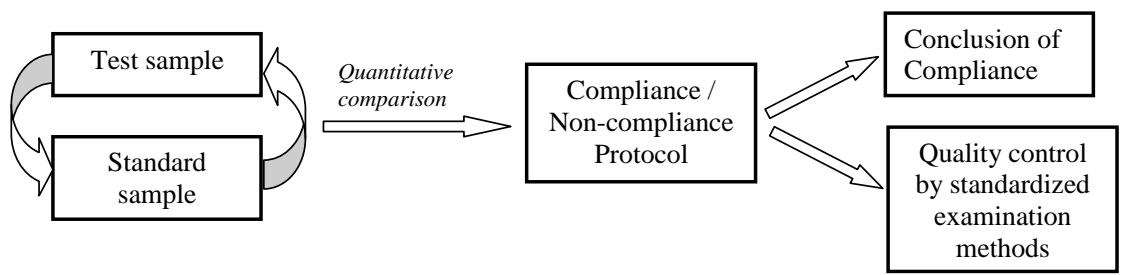

Fig. 10. Scheme of incorporating the proposed method into the procedure of assessing the compliance of a liquid with its standard.

The advantages of the method also include absence of preliminary treatment of samples, comparison of results at room temperatures and moderate humidity, compactness and transportability of the device, its low cost, versatility, ease of use and environmental safety.

**5. Conclusion**

We have presented a new approach to the comparative analysis of liquid food products. A dynamic indicator — an oscillogram of the change in the electrical impedance of the sensor on which a drop of the studied liquid dries — is used as a "fingerprint" of a multicomponent liquid. We have described the design and principle of operation of the developed device, as well as examples of its testing. The proposed approach is efficient for distinguishing samples not only between different types of wine products, but even within each type: cognacs of different varieties, dry and semi-dry red and white wines of different brands, milk of different concentrations. The method is simple, does not require preliminary sample preparation, and works in the range of room temperature and humidity fluctuations. The device has small dimensions and weight and can be used as a compact console for a laptop. The authors believe that the dynamic "fingerprint" can replace the widely used barcode and form the basis for certification of wine products.

Acknowledgements

The authors express deep gratitude to their colleague, Mrs. Nadezhda B. Krivatkina for useful discussion and proof reading the article.

Funding: This research was funded by the Ministry of Education and Science of Russia, grant number 14.Y26.31.0022

Conflicts of Interest: The authors declare no conflict of interest.

All the data can be find in a data repository:

Hard Drinks (https://github.com/TVK-dev/TVK-data/tree/master/hard_drincs)

Wine (https://github.com/TVK-dev/TVK-data/tree/master/wine)
Milk (https://github.com/TVK-dev/TVK-data/tree/master/Milk_database)